%% file: jhelum.tex
\documentclass[twocolumn]{aastex62}

\usepackage[T1]{fontenc}

\newcommand{\acronym}[1]{{\small{#1}}}
\newcommand{\package}[1]{\textsl{#1}}
\newcommand{\gaia}{\textsl{Gaia}}

\shorttitle{the jhelum puzzle}
\shortauthors{bonaca et al.}

\usepackage{amsmath}

\begin{document}\sloppy\sloppypar\raggedbottom\frenchspacing 

\title{Multiple components of the Jhelum stellar stream}

\correspondingauthor{Ana Bonaca}
\email{ana.bonaca@cfa.harvard.edu}

\author[0000-0002-7846-9787]{Ana Bonaca}
\affil{Center for Astrophysics | Harvard \& Smithsonian, 60 Garden St, Cambridge, MA 02138, USA}

\author[0000-0002-1590-8551]{Charlie Conroy}
\affil{Center for Astrophysics | Harvard \& Smithsonian, 60 Garden St, Cambridge, MA 02138, USA}

\author[0000-0003-0872-7098]{Adrian~M.~Price-Whelan}
\affil{Department of Astrophysical Sciences, 4 Ivy Lane, Princeton University, Princeton, NJ 08544, USA}

\author[0000-0003-2866-9403]{David W. Hogg}
\affiliation{Center for Cosmology and Particle Physics,
Department of Physics,
New York University}
\affiliation{Center for Data Science,
New York University}
\affiliation{Max-Planck-Institut f\"ur Astronomie, Heidelberg}
\affiliation{Flatiron Institute, Simons Foundation}

\begin{abstract}\noindent 
In simple models of the Milky Way, tidally disrupting satellites produce long and thin---nearly one-dimensional---stellar streams.
Using astrometric data from the Gaia second data release and photometry from the Dark Energy Survey, we demonstrate that the Jhelum stream, a stellar stream in the inner halo, is a two-dimensional structure.
The spatial distribution of highly probable Jhelum members reveals a dense thin component and an associated diffuse, spatially offset component.
These two spatial components have indistinguishable proper motions (at $\sigma\sim1\,\rm mas\,yr^{-1}$ level) and a similar ratio of blue straggler to blue horizontal branch stars, which indicates a common origin for the two components.
The best-fit orbit to the narrow component (pericenter $8\,\rm kpc$, apocenter $24\,\rm kpc$), however, does not explain the wide component of the Jhelum stream.
On the other hand, an older orbital wrap of Jhelum's orbit traces the Indus stream, indicating a possible connection between these two structures and additional complexity in Jhelum's formation.
Substructure in the Jhelum progenitor or precession of its tidal debris in the Milky Way potential may explain the observed structure of Jhelum.
Future spectroscopic data will enable discrimination between these ``nature'' and ``nurture'' formation scenarios.
Jhelum adds to the growing list of cold stellar streams that display complex morphologies and promise to reveal the dynamical history of the Milky Way.
\end{abstract}

\keywords{%
stars:~kinematics~and~dynamics
  ---
Galaxy:~halo
  ---
Galaxy:~kinematics~and~dynamics
}

\section{Introduction}
\label{sec:intro}
Stars escaping from globular clusters form thin, dynamically cold tidal streams \citep[e.g.,][]{combes1999}.
The phase-space distribution of tidal debris is predominantly determined by the gravitational tidal field, so \citet{johnston1999} proposed measuring the distribution of matter in the Galaxy using stellar streams.
In a time-independent potential, the mean track of a stream constrains the acceleration vector at its current location \citep{bh2018}.
As more than 40 stellar streams have been discovered at a range of distances in the Milky Way halo \citep[see][for a recent review]{gc2016}, streams should provide a three-dimensional map of the Galactic potential.

Being long and thin structures, stellar streams also preserve a historical record of gravitational perturbations on small scales and have been discussed as tracers of dark matter substructure \citep[e.g.,][]{johnston2002, carlberg2009}.
A telltale signature of an interaction with a dark matter subhalo is a gap in the stellar density along the stream \citep[e.g.,][]{ibata2002, yoon2011, eb2015}.
Tantalizing hints of stream gaps were first observed in photometric surveys \citep[e.g.,][]{carlberg2012,cg2013}, and gaps were definitively detected in the GD-1 stellar stream \citep{gd2006} when we used \gaia\ proper motions to cleanly select likely stream members \citep{pwb}.
This discovery opened a new era in which globular cluster streams are no longer simple tracers of the global gravitational potential, but instead provide additional constraints  through their complex internal structure.

In addition to opening gaps along the stream, a dynamic, clumpy and/or time-dependent, environment can disperse stars from originally thin streams to form much wider structures \citep[e.g.,][]{bonaca2014, ngan2016, pw2016, pearson2017}.
Alternatively, a globular cluster that started disrupting in a satellite galaxy before its accretion to the main Milky Way halo, can create a cold stream that is also accompanied by a wide, low surface-brightness component \citep{carlberg2018}.
Wide extensions have not yet been detected around the known cold streams, however, recent improvements in identifying stream members motivate a more comprehensive search.

Here we present the first evidence for two components of the Jhelum stream.
Discovered as a photometric overdensity in the Dark Energy Survey \citep[DES,][]{abbott2018}, Jhelum is a $\sim30^\circ$ long and $\sim1^\circ$ wide stellar stream at a distance of $\sim13\,$kpc \citep{shipp2018}.
Like GD-1, Jhelum is also on a retrograde orbit with respect to the Milky Way disk \citep{malhan2018}, so we use \gaia\ proper motions in addition to DES photometry to better select likely members (\S\ref{sec:data}).
The resulting map of the stream reveals that Jhelum has a thin and a wide component (\S\ref{sec:structure}); we compare and contrast these components in \S\ref{sec:origin}.
In \S\ref{sec:discussion} we conclude with a discussion of possible origin scenarios.

\section{Data}
\label{sec:data}
We start our analysis by defining a coordinate system $(\phi_1,\phi_2)$ that is aligned with the Jhelum stellar stream.
The great circle best-fitting the Jhelum track has a pole $(\alpha_{2000},\delta_{2000}) = (359.1^\circ, 38.2^\circ)$ \citep{shipp2018}.
We use a coordinate system with the origin at $(\alpha,\,\delta) = (359.1^\circ,\,-51.9^\circ)$.
The rotation matrix that converts equatorial $(\alpha, \delta)$ coordinates to Jhelum coordinates $(\phi_1, \phi_2)$, where $\phi_1$ is the coordinate along the stream and $\phi_2$ is perpendicular to the stream track, is available electronically at \url{https://github.com/abonaca/jhelum} and is implemented as a stream coordinate frame in \texttt{Gala} \citep{gala}.
In these coordinates, Jhelum is centered at $\phi_2=0$, and the DES detection spans $-5^\circ\lesssim\phi_1\lesssim25^\circ$ \citep{shipp2018}.

\begin{figure}
\begin{center}
\includegraphics[width=0.99\columnwidth]{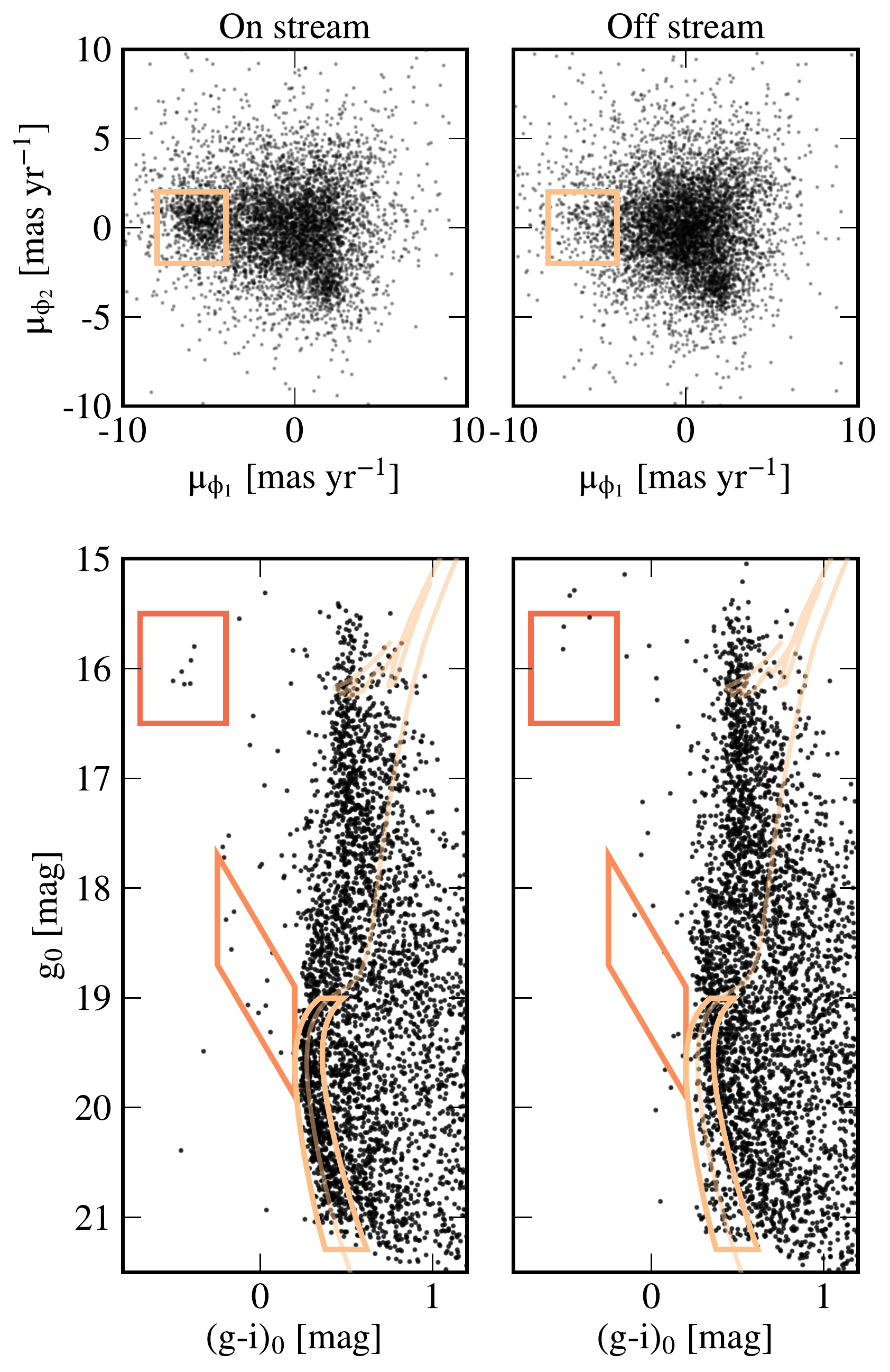}
\end{center}
\caption{
(Top) Proper motions of photometry-selected stars along the Jhelum stream (left) and in a control field (right).
(Bottom) Similarly, color-magnitude diagrams of stars selected on proper motions.
Photometric and proper-motion selection boxes are shown in light orange.
Jhelum stands out from the Milky Way field population in proper motions as a retrograde stream, and in the color-magnitude diagram where its main sequence is more metal-poor than the field, and is accompanied by blue stragglers (medium orange) and blue horizontal branch stars (dark orange box).
}
\label{fig:properties}
\end{figure}

We query the Gaia DR2 \citep{gdr2} and DES DR1 \citep{abbott2018} catalogs between $-10^\circ<\phi_1<35^\circ$ and $-5^\circ<\phi_2<5^\circ$, and select all sources identified as stars that are brighter than $g_0<23$, while excluding stars with parallaxes larger than $1\,\rm mas$.
DES photometry was dereddened using \citet{sfd} dust maps.
Assuming a constant distance along the stream of 13\,kpc, we correct the whole catalog for the solar reflex motion following \citet{pwb}.

\begin{figure*}
\begin{center}
\includegraphics[width=0.9\textwidth]{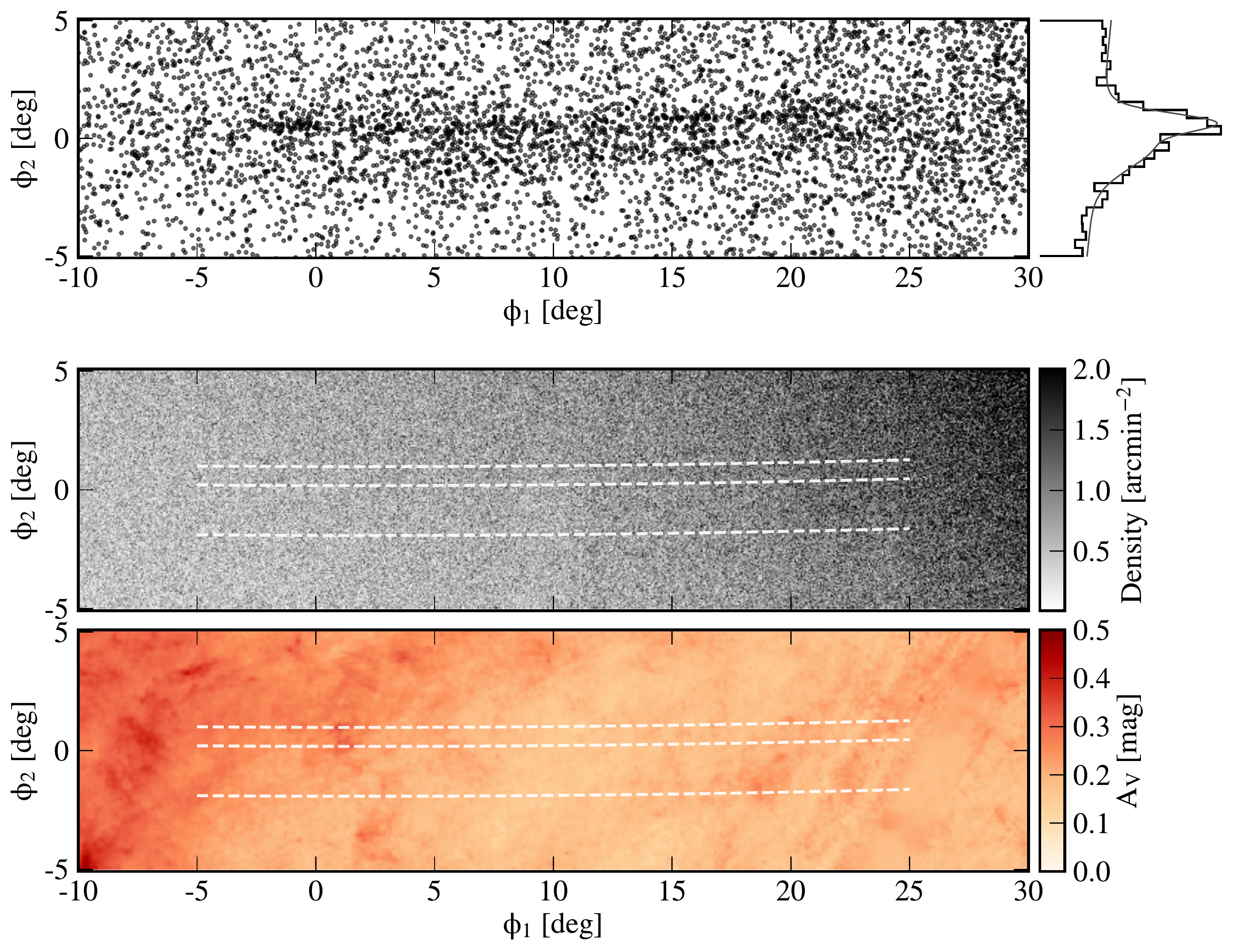}
\end{center}
\caption{
(Top left) Sky positions of likely Jhelum members in the stream coordinate system reveal for the first time a complex morphology in a cold stream.
Member selection is based on the \gaia\ proper motions and DES photometry, and excludes nearby contaminants using \gaia\ parallaxes.
(Top right) Profile of likely Jhelum members between $-5^\circ<\phi_1<25^\circ$ is asymmetric, with a narrow, dense component at $\phi_2>0^\circ$ and a more diffuse, wide component at $\phi_2<0^\circ$.
This morphology is intrinsic to the stellar stream, as similar signatures are absent from the full stellar density field (middle) and the dust map (bottom).
}
\label{fig:map}
\end{figure*}

Following these corrections, Jhelum stars are clearly separated from the Milky Way field population in the proper motion and color-magnitude spaces.
Figure~\ref{fig:properties} shows proper motions (top) and color-magnitude diagram (bottom) for a stream field ($0^\circ<\phi_1<25^\circ$, $0^\circ<\phi_2<1^\circ$, left) and a comparison field ($0^\circ<\phi_1<25^\circ$, $3.5^\circ<|\phi_2|<4^\circ$, right).
In proper motions, Jhelum stands out from the Milky Way as a retrograde stream, and we select likely members between $-8<\mu_{\phi_1,\star}/{\rm mas\,yr^{-1}}<-4$ and $-2<\mu_{\phi_2}/{\rm mas\,yr^{-1}}<2$.
The stream also appears as a prominent overdensity of main sequence stars, which we select following a 12\,Gyr old, metal-poor ($\rm[Fe/H]=-1.5$) MIST isochrone \citep{choi2016} between $19<g_0<21.3$.
Both of these selection regions are shown in light orange in Figure~\ref{fig:properties}.

\section{Density structure of Jhelum}
\label{sec:structure}

\begin{figure*}
\begin{center}
\includegraphics[width=0.9\textwidth]{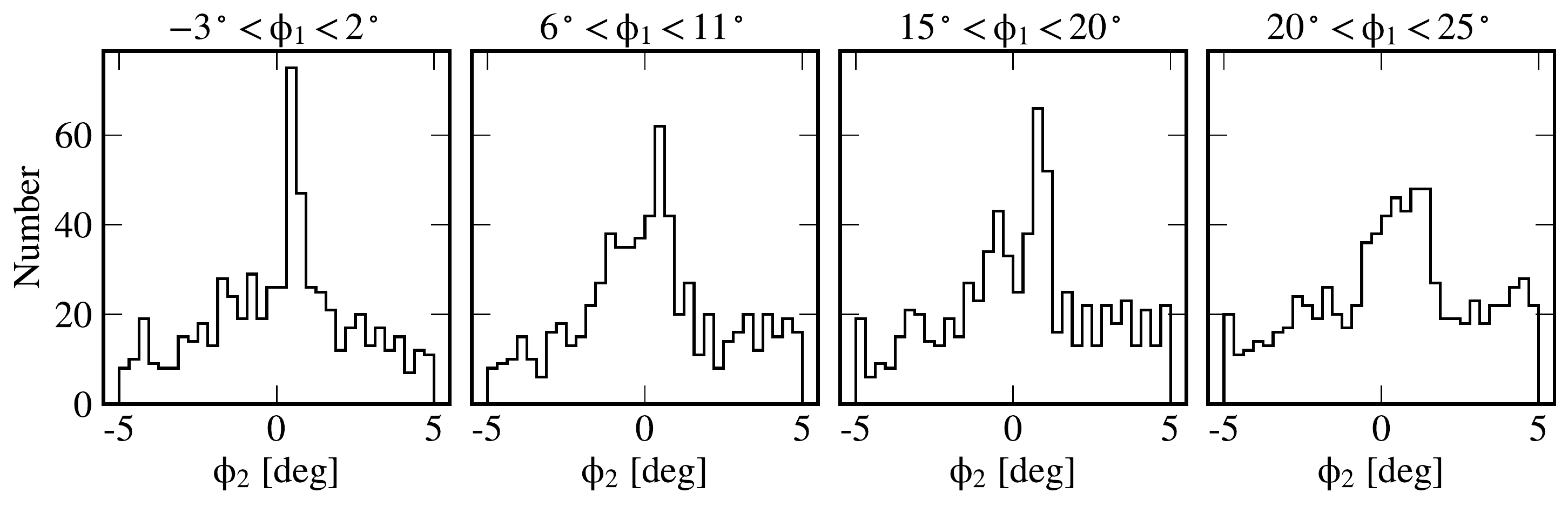}
\end{center}
\caption{
The profile of likely Jhelum members changes along the stream.
Panels show $5^\circ$ wide regions along the stream, with the location in $\phi_1$ (indicated at the top) growing from left to right.
Between $-3^\circ<\phi_1<2^\circ$, only a narrow component of the stream is present, while in the next $\phi_1$ bin, $6^\circ<\phi_1<11^\circ$, the stream features both the narrow and the wide component.
Further along the stream, $15^\circ<\phi_1<20^\circ$, the stream consists of two narrow components separated by a gap, while at the trailing end, $20^\circ<\phi_1<25^\circ$, the gap vanishes and the stream appears as a single, wide feature.
}
\label{fig:histo}
\end{figure*}

Sky positions of Jhelum members selected in Section~\ref{sec:data} are presented in the top left panel of Figure~\ref{fig:map}.
Although some contamination from the Milky Way field stars remains, the stream stands out as an overdensity between $-5^\circ<\phi_1<25^\circ$, $-1^\circ<\phi_2<1^\circ$.
Despite the increase in the purity of the stream membership, this extent is similar to the initial detection reported by \citet{shipp2018}.
However, the new data reveal unexpected internal structure of the stream: the density of stream members is higher at $\phi_2>0$ than at $\phi_2<0$.
The $\phi_2$ distribution of likely Jhelum members (Figure~\ref{fig:map}, top right) has two clear components, with a more prominent, narrow component at $\phi_2>0$, and a less prominent, diffuse component peaking at $\phi_2\approx0$.

Surveys such as Gaia and DES have complex selection functions \citep[e.g.,][]{bovy2017}, which can imprint density inhomogeneities in stellar maps.
To test whether the density structure observed in Jhelum is inherited from a survey strategy, we show a density map of all stars in our input catalog (Gaia crossmatched with DES, and with parallax $\varpi<1\,\rm mas$) in the middle panel of Figure~\ref{fig:map}.
Dashed white lines bracket the two Jhelum components.
The lines are offset from the best-fitting polynomial to the running median of the dense Jhelum component:
\begin{equation}
\input{polytrack}
\label{eq:track}
\end{equation}
where $\phi_1$ and $\phi_2$ are in degrees.
While there is a large density gradient along the $\phi_1$ direction, as positive $\phi_1$ values correspond to lower galactic latitudes, the overall stellar density changes little across the stream in the $\phi_2$ direction at a fixed $\phi_1$ location.

Density variations observed in streams can also originate from nonuniform dust attenuation \citep[e.g.,][]{ibata2016}.
In that case, the features observed in the stream correlate with a dust map.
Extinction along Jhelum varies between $A_V\sim0.2-0.5$ \citep[Figure~\ref{fig:map}, bottom;][]{sfd}.
The regions of high dust attenuation at $(\phi_1,\phi_2)\approx(1^\circ,0.5^\circ)$ and $(\phi_1,\phi_2)\approx(19^\circ,-2^\circ)$ correspond to regions of reduced Jhelum density, however, there are no global gradients in dust extinction perpendicular to the stream.
Therefore, we conclude that the transverse variations in Jhelum density are intrinsic to the stream itself.

To quantify substructure in the Jhelum stellar stream, we model the $\phi_2$ distribution of likely stream members (Figure~\ref{fig:map}, top right).
We assume a mixture model with two Gaussian components (defined by means $\mu_{1,2}$ and variances $\sigma^2_{1,2}$ for the narrow and wide component, respectively) and a background which is allowed to linearly vary with $\phi_2$ (defined by the gradient $a\rm_{bg}$).
The density model for a given set of parameters $\theta = (\alpha_1, \alpha_2, \alpha_{\rm bg}, \mu_1, \mu_2, \sigma_1, \sigma_2, a_{\rm bg})$ is:
\begin{equation}
\begin{aligned}
p(\phi_2 | \theta) = &\alpha_1 \mathcal{N}(\phi_2 | \mu_1, \sigma_1) + \alpha_2 \mathcal{N}(\phi_2 | \mu_2, \sigma_2) \\
&+\alpha_{\rm bg} (a_{\rm bg}\phi_2 + \mathcal{U}(-5,5))
\label{eq:model}
\end{aligned}
\end{equation}
where $\mathcal{N}$ and $\mathcal{U}$ are the normal and uniform distributions, $\alpha_{1,2}$ are the fractions of stars in the narrow and wide components, respectively, and $\alpha_{\rm bg} = 1 - \alpha_1 - \alpha_2$ is the fraction of the Milky Way field stars.
We sample the parameter space $\theta$ with an affine invariant Markov Chain Monte Carlo ensemble sampler \citep{emcee}.
The sampler ran with 64 walkers for 4096 steps, the first half of which were discarded as the burn-in, assuming flat priors in normalizations and means, and log-uniform priors for the variances and the background gradient.
The highest-likelihood model (gray line in the top right of Figure~\ref{fig:map}) reproduces well the observed distribution of Jhelum stars.
The amplitudes of both Gaussian components are statistically significant: \input{a1} for the narrow, and \input{a2} for the diffuse component.
Their respective widths are \input{s1}\,deg and \input{s2}\,deg.
At a distance of 13\,kpc, this corresponds to \input{w1}\,pc and \input{w2}\,pc, respectively, which is comparable to the widths of known globular cluster streams in the Milky Way (e.g., Palomar~5 stream is $\sim120$\,pc wide, \citealt{odenkirchen2003}).

The prominence of the two Jhelum components changes along the stream (see Figure~\ref{fig:map}).
In Figure~\ref{fig:histo} we show the transverse density profiles in four non-overlapping $\phi_1$ regions.
Splitting the original sample increases the influence of Poisson statistics in the profiles, so we only discuss their features qualitatively.
The left-most panel shows that the leading end of the stream ($-3^\circ<\phi_1<2^\circ$) consists of a single, narrow component.
Two components are detected both between $6^\circ<\phi_1<11^\circ$ and between $15^\circ<\phi_1<20^\circ$, however, while the component at smaller $\phi_2$ is wide in the former region (second panel from left), it is narrow in the latter and clearly separated from the narrow component at larger $\phi_2$ (third panel).
At Jhelum's trailing end ($20^\circ<\phi_1<25^\circ$, right-most panel), there is again a single component, but almost twice as wide as that on the leading end.
Curiously, the narrow component visible in the first three panels has an approximately constant width of $\sim0.25^\circ$ ($\sim60\,$pc).
This diversity of transverse density profiles along Jhelum underlines the intricacy of its formation history.

\section{Properties of the Jhelum components}
\label{sec:origin}
The Jhelum stellar stream appears to have two spatially distinct components (Figure~\ref{fig:map}).
In this section we compare structural and dynamical properties of these components to uncover their origin.

\begin{figure}
\begin{center}
\includegraphics[width=\columnwidth]{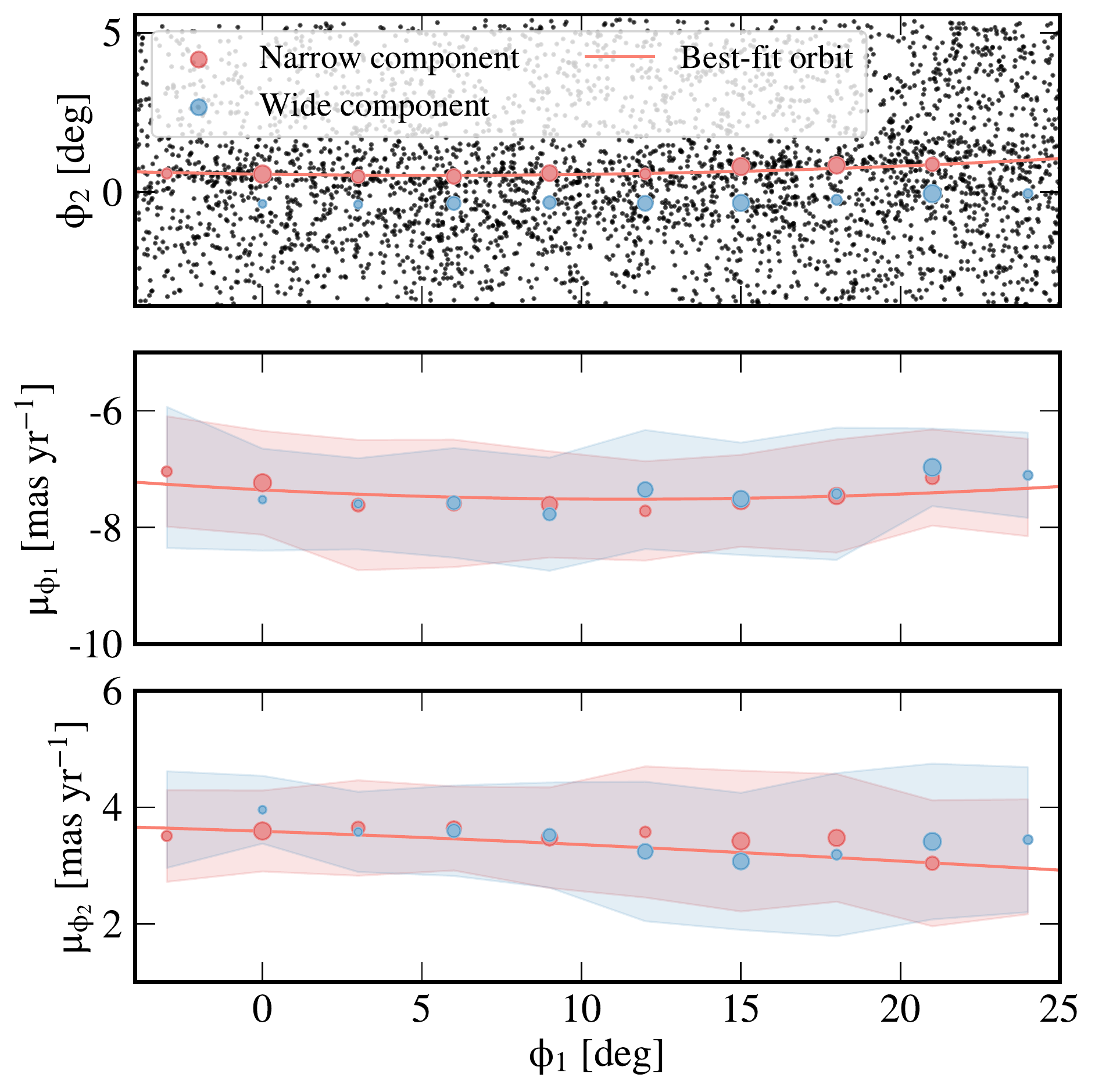}
\end{center}
\caption{
On-sky distribution (top) and proper motion profiles (middle, bottom) of the narrow and wide Jhelum components (red and blue points, respectively).
Spatially offset by $\sim0.9^\circ$, the two Jhelum components have proper motions consistent within uncertainties (shaded area).
The best-fitting orbit in the standard Milky Way potential reproduces the sky and proper motion distribution of the narrow component (thin line).
}
\label{fig:components}
\end{figure}

We first analyze the stellar population of Jhelum.
Interestingly, blue horizontal branch (BHB) and blue straggler (BS) stars are present in Jhelum, and they are hardly contaminated with field stars (bottom panels of Figure~\ref{fig:properties}).
The ratio of BHB to BS stars can distinguish between massive dwarf galaxy and globular cluster progenitors \citep[e.g.,][]{momany2007, deason2015}, so we characterize the Jhelum stellar population with the BS to BHB ratio.
We select BHBs at the Jhelum distance with: $-0.7<g-i<-0.2$, $15.5<g<16.5$ (dark orange box in Figure~\ref{fig:properties}) and BSs within the polygon $(g-i,g) = [(0.2,18.9), (0.2, 19.9), (-0.25, 18.7), (-0.25,17.7)]$ (medium orange box).
In the Jhelum footprint (both spatial and proper-motion) there are a total of 31 BSs and 12 BHBs, compared to 12 BS and 2 BHB stars in the control field.
Subtracting the field population yields an intrinsic BS to BHB ratio of $N_{BS} / N_{BHB} = 1.9\pm 0.7$
This ratio is consistent with a dwarf galaxy progenitor in a wide mass range ($M_V\approx-6$~to~$-11$), as well as with a low-mass, $M_V\lesssim-6$, globular cluster progenitor \citep{deason2015}.
Split between the two components, the ratio becomes $1.7\pm0.9$ and $2.1\pm1.2$ for the narrow and wide component, respectively.
Within uncertainties, the BHB to BS ratio is the same in the two Jhelum components, indicating a common origin.
However, detailed chemical abundances from spectroscopy are required to definitively establish the single-progenitor scenario.

Next, we map Jhelum components in the sky and in proper motions.
We spatially define the narrow component as stars within $0.4^\circ$ from the ridgeline (Equation~\ref{eq:track}), and the wide component as stars $0.4^\circ-2^\circ$ below the ridgeline (as shown in Figure~\ref{fig:map}).
For each component, we calculate average properties in $4^\circ$-wide, overlapping bins spaced by $3^\circ$ in the $\phi_1$ direction.
From top to bottom, Figure~\ref{fig:components} shows the running medians of $\phi_2$ positions, $\mu_{\phi_1}$ and $\mu_{\phi_2}$ proper motion components, with red and blue points for the narrow and wide components, respectively.
The size of the point scales with the number of Jhelum stars in the bin, while the shaded area encompasses the median absolute deviation in each component's track.

Both Jhelum components trace a great circle, with their tracks only slightly curving from the $\phi_2=0$ line (Figure~\ref{fig:components}, top).
The components are offset by $\sim0.9^\circ$ in the $\phi_2$ direction, and the offset is constant along the stream.
Despite being spatially offset, the proper motions of the two components are remarkably similar (Figure~\ref{fig:components}, middle and bottom).
The dispersion in proper motions is large ($0.7-1.2\,\rm mas\,yr^{-1}$) and comparable to the observational uncertainties (median for likely Jhelum members is $0.7\rm\, mas\, yr^{-1}$), both of which are much smaller than the typical kinematic offset between the two components ($\lesssim0.3\,\rm mas\,yr^{-1}$).
At the current precision, the Jhelum components are kinematically indistinguishable.

\begin{figure}
\begin{center}
\includegraphics[width=\columnwidth]{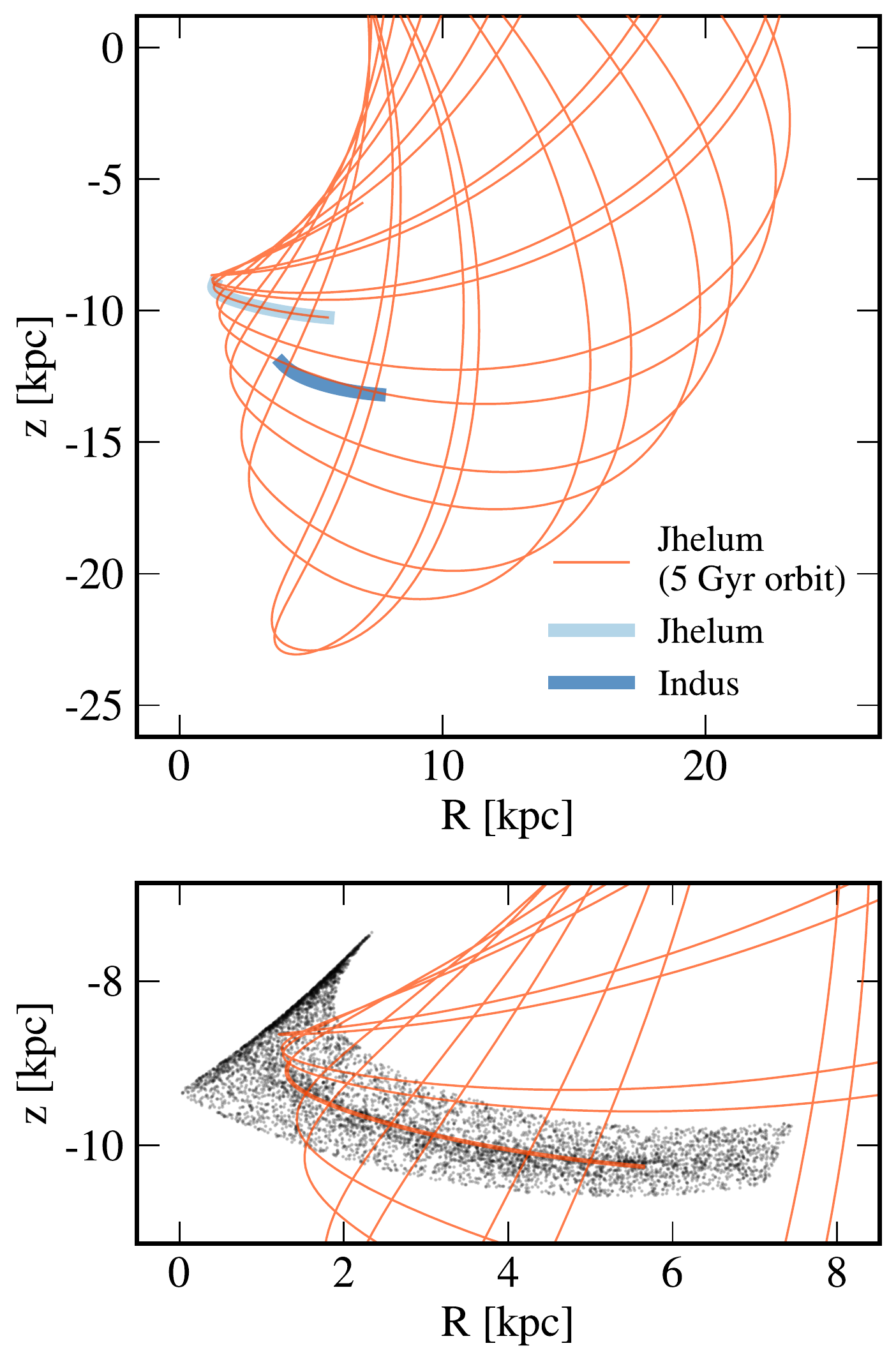}
\end{center}
\caption{
Best-fitting orbit of the Jhelum stellar stream in cylindrical Galactocentric coordinates (top).
The thick light blue line denotes the observed extent of Jhelum.
The same orbit simultaneously matches the location of the Indus stream (thick dark blue line).
The dense Jhelum component is closer to the Galactic plane than the diffuse component (bottom).
}
\label{fig:galactocentric}
\end{figure}

Finally, we explore whether both Jhelum components can be explained within a simple dynamical model.
Assuming a standard Milky Way potential \citep{gala}, we use the BFGS minimization algorithm to search for orbits that simultaneously fit the sky distribution of the narrow component, its proper motions and that place the stream at a constant distance of 13\,kpc (similar to the orbit-fitting method defined and used in \citealt{pwb}).
The best-fitting orbit, shown as a thin line in Figure~\ref{fig:components}, matches the observed track and proper motion gradients.

In Figure~\ref{fig:galactocentric} we present the Jhelum stream and its best-fitting orbit in cylindrical Galactocentric coordinates.
The thin orange line traces Jhelum's orbit for the last 5\,Gyr, while the thick light blue line marks the present-day extent of the stream (top panel).
Jhelum orbits between 8 and 24\,kpc, with a period of $\sim300$\,Myr, and is currently just past pericenter.

Remarkably, a past orbital wrap of the Jhelum stream traces the Indus stream in the Galactocentric $R-z$ plane (thick dark blue line in top of Figure~\ref{fig:galactocentric}, based on the sky positions and typical distance reported in \citealt{shipp2018}).
This agreement may be dynamical evidence that Indus and Jhelum are different orbital arms of the same progenitor, first suggested by \citet{shipp2018} based on the streams' similar width, $\sim1^\circ$, and photometric metallicity, $\rm[Fe/H]\sim-1.4$, as well as their physical proximity.
A fairly massive progenitor would be required to produce the combined stellar debris of Indus and Jhelum, and the high metallicity as well as the high ratio of BS to BHB stars indeed indicate a massive progenitor.
Furthermore, both streams may extend beyond their currently measured extent.
For example, Figure~\ref{fig:map} shows that our detection of Jhelum is impacted by the Milky Way disk at $\phi_1\gtrsim-25^\circ$ and by dust at $\phi_1\lesssim-5^\circ$.
If the connection is confirmed, Indus and Jhelum would be one of the longest tidal structures in the Milky Way, and therefore extremely constraining for its gravitational potential \citep{bh2018}.
A similar analysis of Indus is required to definitively establish the two streams are dynamically related, which we defer to future work.

The bottom panel of Figure~\ref{fig:galactocentric} zooms in on Jhelum's current location, and also shows the distribution of its likely members (assuming the distance gradient from the best-fit orbit).
Jhelum's wide component is $\sim0.5\,\rm kpc$ further from the Galactic plane than its narrow component.
The orbit matches the narrow Jhelum component, but none of the previous orbital wraps pass through the wide component, arguing against the two components being debris from a single progenitor released at subsequent pericentric passages.

\section{Discussion}
\label{sec:discussion}
The combination of Gaia astrometry and DES photometry allows us to cleanly select members of the Jhelum stream.
The resulting map reveals Jhelum has two statistically-significant parallel components, separated by $\sim0.9^\circ$ ($\sim200\,\rm pc$).
The two components have similar stellar populations and similar proper motion gradients, which suggests they originate from the same progenitor.
Current constraints on the stream's metallicity and relative abundance of blue straggler and blue horizontal branch stars cannot distinguish between a globular cluster and a dwarf galaxy progenitor.

Here we discuss possible origin scenarios for the two components of the Jhelum stream in the context of these observations:
\begin{itemize}
 \item{\emph{Multiple progenitors:} the similar ratios of blue straggler to blue horizontal branch stars suggest that the two components have the same stellar population, but there is still room for two distinct progenitors that have similar BS to BHB ratios \citep[e.g., a system of a low-mass globular cluster and a low-mass dwarf galaxy,][]{deason2015}.
 This hypothesis can be directly tested by measuring chemical abundances in the two components.
 }
 \item{\emph{Kinematic substructure in the progenitor:} substructure in the progenitor may lead to non-trivial density structure of its tidal debris.
 For example, two spatially distinct components can form if the progenitor is a globular cluster that is initially orbiting in a dark-matter subhalo \citep[e.g.,][]{penarrubia2017, carlberg2018}.
 In this scenario, the wide component originates from stars stripped while the globular cluster was still in the subhalo \citep[similar to the recently reported GD-1 cocoon,][]{malhan2019}, while the narrow component would be stars more recently released directly in the Milky Way gravitational potential.
 The velocity dispersion in each component should reflect their local environment prior to the formation of the stream \citep[e.g.,][]{fardal2015}.
 Precise measurements of Jhelum's proper motions or radial velocities can test whether the wide component is indeed kinematically hotter than the narrow one.
 }
 \item{\emph{Different orbital wraps:} similar to Indus being aligned with an old orbital wrap of the Jhelum's orbit, the Jhelum components may originate from different orbital passages of a single progenitor.
 While our best-fit orbit does not simultaneously pass through both components, the Galactocentric $z$ separation of several-Gyr-old orbital wraps is similar to that of the Jhelum components (Figure~\ref{fig:galactocentric}).
 Better characterization of the orbit through more precise measurements of the stream distance and kinematics will test this scenario.
 If Jhelum's wide component is indeed an old wrap of the orbit that best-fits its narrow component, the stream will put extremely strong constraints on the gravitational potential.
 }
 \item{\emph{Fold caustic:} tidal debris distributed in a plane, but viewed almost edge-on, could produce the density profile observed in Jhelum.
 Two-dimensional shells are commonly observed \citep[e.g.,][]{tal2009,kadofong2018}, however, their densest part, unlike Jhelum's, is at the largest galactocentric radius.
 A more general fold caustic of a fully phase-mixed distribution is still allowed \citep[e.g.,][]{tremaine1999}, in which case the velocity dispersion in the dense component of Jhelum should be higher than in its diffuse part.
 Precise kinematics will test this formation pathway as well.
 }
 \item{\emph{Precession of the orbital plane:} streams orbiting in non-spherical potentials widen because the stream star orbits differentially precess \citep[e.g.,][]{erkal2016, dehnen2018}.
 Jhelum's orbit is significantly affected by the Milky Way disk, so its extended structure may be attributed to differential orbital precession.
 However, the expected width of a stream on Jhelum's orbit in the fiducial Milky Way potential is only a fraction of the observed width.
 Jhelum models in more asymmetric potentials need to be explored to test this scenario.
 }
 \item{\emph{Chaos:} streams formed on chaotic (even weakly-chaotic) orbits may develop low surface-brightness envelopes \citep[e.g.,][]{pw2016}.
 However, in our simple gravitational potential, Jhelum's orbit is regular (see Figure~\ref{fig:galactocentric}):
 Within the Galactocentric radii relevant to Jhelum, the global mass distribution is likely close to spherical or mildly oblate \citep[e.g.,][]{kupper2015}, and thus chaos driven by the global potential is likely not relevant for Jhelum.
 }
 \item{\emph{Time-dependent perturbations:} massive, dynamical perturbers such as the rotating bar or Large Magellanic Cloud (LMC) can affect the structure of stellar streams \citep[e.g.,][]{pw2016b, pearson2017, erkal2019}.
 Jhelum is on a retrograde orbit in the inner Galaxy, which limits the influence of both the LMC and the bar.
 However, perturbations from a population of low-mass objects can also result in complex morphologies of stellar streams \citep[e.g.,][]{bonaca2014}, and remain a viable mechanism for shaping the Jhelum stream.
 }
\end{itemize}

All of these formation scenarios merit further investigation, but solutions where Jhelum remains a coherent tidal structure on a largely unperturbed orbit appear more likely.
Our best-fit orbit for Jhelum simultaneously (and independently) matches the Indus stream, suggesting that only minor perturbations are allowed from the bar, chaos or LMC.
Both streams are still coherent, so this argues against the fold caustic interpretation for Jhelum's vertical structure.

A combination of spectroscopic data and more detailed dynamical modeling can further constrain Jhelum's formation scenario.
Chemical abundances will determine whether both Jhelum components originate from the same progenitor, as well as distinguish between a globular cluster and a dwarf galaxy origin.
In the case of a single progenitor, confronting the theoretical and observed radial velocities in the two Jhelum components will differentiate between them being different substructures within the progenitor or differentially precessing debris.

The transverse structure that \gaia\ revealed in the Jhelum stream is evidence of a formation mechanism beyond simple tidal disruption.
The only other stream studied to a similar level of detail with \gaia\ is GD-1 \citep{pwb}, where the discovered off-stream features may be evidence of a recent perturbation \citep{bonaca2018}.
These discoveries signal the dawn of a new era, in which the internal structure of thin stellar streams is used to trace the structure of their formative environment and the Galaxy.

\vspace{0.5cm}
\acknowledgements
We thank Ray Carlberg, Denis Erkal, Kathryn Johnston, Nora Shipp, Josh Speagle, Scott Tremaine for helpful conversations.

This project was developed in part at the 2018 \acronym{NYC} \gaia\ \acronym{DR2} Workshop at the Center for Computational Astrophysics of the Flatiron Institute in New York City in 2018 April.

This work has made use of data from the European Space Agency (\acronym{ESA}) mission \gaia\ (\url{https://www.cosmos.esa.int/gaia}), processed by the \gaia\ Data Processing and Analysis Consortium (\acronym{DPAC}, \url{https://www.cosmos.esa.int/web/gaia/dpac/consortium}). Funding for the \acronym{DPAC} has been provided by national institutions, in particular the institutions participating in the \gaia\ Multilateral Agreement.

This project used public archival data from the Dark Energy Survey (DES). Funding for the DES Projects has been provided by the U.S. Department of Energy, the U.S. National Science Foundation, the Ministry of Science and Education of Spain, the Science and Technology FacilitiesCouncil of the United Kingdom, the Higher Education Funding Council for England, the National Center for Supercomputing Applications at the University of Illinois at Urbana-Champaign, the Kavli Institute of Cosmological Physics at the University of Chicago, the Center for Cosmology and Astro-Particle Physics at the Ohio State University, the Mitchell Institute for Fundamental Physics and Astronomy at Texas A\&M University, Financiadora de Estudos e Projetos, Funda{\c c}{\~a}o Carlos Chagas Filho de Amparo {\`a} Pesquisa do Estado do Rio de Janeiro, Conselho Nacional de Desenvolvimento Cient{\'i}fico e Tecnol{\'o}gico and the Minist{\'e}rio da Ci{\^e}ncia, Tecnologia e Inova{\c c}{\~a}o, the Deutsche Forschungsgemeinschaft, and the Collaborating Institutions in the Dark Energy Survey.

The Collaborating Institutions are Argonne National Laboratory, the University of California at Santa Cruz, the University of Cambridge, Centro de Investigaciones Energ{\'e}ticas, Medioambientales y Tecnol{\'o}gicas-Madrid, the University of Chicago, University College London, the DES-Brazil Consortium, the University of Edinburgh, the Eidgen{\"o}ssische Technische Hochschule (ETH) Z{\"u}rich,  Fermi National Accelerator Laboratory, the University of Illinois at Urbana-Champaign, the Institut de Ci{\`e}ncies de l'Espai (IEEC/CSIC), the Institut de F{\'i}sica d'Altes Energies, Lawrence Berkeley National Laboratory, the Ludwig-Maximilians Universit{\"a}t M{\"u}nchen and the associated Excellence Cluster Universe, the University of Michigan, the National Optical Astronomy Observatory, the University of Nottingham, The Ohio State University, the OzDES Membership Consortium, the University of Pennsylvania, the University of Portsmouth, SLAC National Accelerator Laboratory, Stanford University, the University of Sussex, and Texas A\&M University.

Based in part on observations at Cerro Tololo Inter-American Observatory, National Optical Astronomy Observatory, which is operated by the Association of Universities for Research in Astronomy (AURA) under a cooperative agreement with the National Science Foundation.


\software{
    \package{Astropy} \citep{astropy, astropy2018},
    \package{gala} \citep{gala},
    \package{matplotlib} \citep{mpl},
    \package{numpy} \citep{numpy},
    \package{scipy} \citep{scipy}
}

\bibliographystyle{aasjournal}
\bibliography{jhelum}

\end{document}

%% file: polytrack.tex
\phi_2 = 0.000546 \,\phi_1^2 -0.00217 \,\phi_1 +0.583

%% file: a1.tex
$\alpha_{1} = 0.102^{+0.008}_{-0.024}$

%% file: a2.tex
$\alpha_{2} = 0.161^{+0.010}_{-0.022}$

%% file: s1.tex
$\sigma_{1} = 0.40^{+0.02}_{-0.06}$

%% file: s2.tex
$\sigma_{2} = 0.94^{+0.04}_{-0.10}$

%% file: w1.tex
$w_{1} = 91^{+4}_{-13}$

%% file: w2.tex
$w_{2} = 213^{+8}_{-23}$